\documentclass[aps,prd,superscriptaddress,twocolumn]{revtex4-1}
\usepackage{amssymb}
\usepackage{amsmath}
\usepackage{wallpaper}
\usepackage{bm}
\usepackage{multirow}
\usepackage[colorlinks,linkcolor=blue, urlcolor=blue, anchorcolor=blue, citecolor=blue]{hyperref}

\begin{document}

\title{Mixed phases of compact star matter in a unified mean-field approach} 

\author{Cheng-Jun Xia}
\email{cjxia@yzu.edu.cn}
\affiliation{Center for Gravitation and Cosmology, College of Physical Science and Technology, Yangzhou University, Yangzhou 225009, China}

\author{Toshiki Maruyama}
\email{maruyama.toshiki@jaea.go.jp}
\affiliation{Advanced Science Research Center, Japan Atomic Energy Agency, Shirakata 2-4, Tokai, Ibaraki 319-1195, Japan}

\author{Nobutoshi Yasutake}
\email{nobutoshi.yasutake@it-chiba.ac.jp}
\affiliation{Department of Physics, Chiba Institute of Technology (CIT), 2-1-1 Shibazono, Narashino, Chiba, 275-0023, Japan}
\affiliation{Advanced Science Research Center, Japan Atomic Energy Agency, Shirakata 2-4, Tokai, Ibaraki 319-1195, Japan}

\author{Toshitaka Tatsumi}
\email{tatsumitoshitaka@gmail.com}
\affiliation{Kitashirakawa Kamiikeda-Cho 52-4, Kyoto 606-8287, Japan}

\date{\today}

\begin{abstract}
Based on an extended NJL model that treats baryons as clusters of quarks, we investigate the properties and microscopic structures of mixed phases for various types of first-order phase transitions in a unified manner, where the model parameters are fixed by reproducing nuclear matter properties and the binding energies of finite nuclei. In particular, based on the Thomas-Fermi approximation, we investigate the mixed phases arise from the liquid-gas phase transition of nuclear matter, chiral phase transition, and deconfinement phase transition in dense stellar matter adopting spherical and cylindrical approximations for the Wigner-Seitz cells. It is found that the geometrical structures do not emerge for chiral phases transition, while the droplet, rod, slab, tube, and bubble phases emerge sequentially as density increases for the liquid-gas and deconfinement phase transitions. Additional attractive interactions between strange quark matter and hyperons are observed as the deconfinement phase transition is entangled with chiral phase transition of $s$ quarks. The results obtained here should be useful to understand the properties and structures of dense stellar matter throughout compact stars and in particular the matter state in the core regions. Meanwhile, more extensive investigations in a three-dimensional geometry with large box sizes are necessary for our future study.
\end{abstract}

\maketitle

\section{\label{sec:intro}Introduction}

Pulsar-like objects are considered as compact stars with masses ranging from $\sim$1 $M_\odot$ to $\sim$2 $M_\odot$ and radii $\sim$10 km. The average density of such compact stars is thus several times the nuclear saturation density $n_0$ ($\approx 0.16\ \mathrm{fm}^{-3}$), while the density from the surface to center varies many orders of magnitude due to the strong gravitational force. Throughout the vast density range of compact stars, there could be several types of first-order phase transitions, indicating the existence of mixed phases with various possible microscopic structures. The existence of those nonuniform structures could in principle affect the properties and evolutions of compact stars.

At subsaturation densities, nuclear matter exhibits a first-order liquid-gas phase transition, which forms various nonuniform structures and are typically found in the crusts of neutron stars. In particular, due to the strong Coulomb repulsion, nuclei may be deformed and take exotic shapes resembling pasta, i.e., nuclear pasta~\cite{Baym1971_ApJ170-299, Negele1973_NPA207-298, Ravenhall1983_PRL50-2066, Hashimoto1984_PTP71-320, Williams1985_NPA435-844}. Employing spherical and cylindrical approximations of the Wigner-Seitz (WS) cell~\cite{Pethick1998_PLB427-7, Oyamatsu1993_NPA561-431, Maruyama2005_PRC72-015802, Avancini2008_PRC78-015802, Avancini2009_PRC79-035804, Shen2011_ApJ197-20, Gupta2013_PRC87-028801, Togashi2017_NPA961-78}, it was shown that there exist at least five types of geometrical structures, i.e, droplets, rods, slabs, tubes and bubbles, which normally appear sequentially as we increase the density. Further investigations have revealed more complicated structures~\cite{Magierski2002_PRC65-045804, Newton2009_PRC79-055801, Fattoyev2017_PRC95-055804}, e.g., gyroid and double-diamond morphologies~\cite{Nakazato2009_PRL103-132501, Schuetrumpf2015_PRC91-025801}, P-surface configurations~\cite{Schuetrumpf2013_PRC87-055805, Schuetrumpf2019_PRC100-045806}, nuclear waffles~\cite{Schneider2014_PRC90-055805, Sagert2016_PRC93-055801}, Parking-garage structures~\cite{Berry2016_PRC94-055801}, deformed droplets~\cite{Kashiwaba2020_PRC101-045804}, as well as the mixtures of droplets and rods, slabs and tubes~\cite{Watanabe2003_PRC68-035806, Okamoto2012_PLB713-284}. It was demonstrated that the structures and properties of those nuclear pastas play important roles in the transport and elastic properties of neutron stars~\cite{Chamel2008_LRR11-10, Caplan2017_RMP89-041002, Zhu2023_PRD107-83023, Sotani2024_Universe10-231}.

Beside nucleons, various exotic hadrons may emerge inside compact stars at slightly larger densities, which may lead to first-order phase transitions as well, e.g., the emergence of pions~\cite{Muto1993_PTPS112-221, Akmal1998_PRC58-1804}, kaons~\cite{Maruyama1994_PLB337-19, Muto2017_JPSCP14-020809}, $\Delta$  resonances~\cite{Sun2019_PRD99-023004}, and hyperons~\cite{Schaffner-Bielich2002_PRL89-171101}. Consequently, the mixed phases of those dense hadronic matter may exist inside compact stars, forming various geometrical structures~\cite{Maruyama2006_PRC73-035802}. At even larger densities, due to the asymptotic freedom of strong interaction, a deconfinement phase transition is expected to take place and transform hadronic matter into quark matter. Nevertheless, so far it is still unclear if the deconfinement phase transition is of first-order or a smooth crossover~\cite{Fukushima2005_PRD71-034002, Voskresensky2023_PPNP130-104030}. If a first-order deconfinement phase transition takes place, then there exist quark-hadron mixed phases~\cite{Glendenning2000, Peng2008_PRC77-065807, Li2015_PRC91-035803, Klahn2013_PRD88-085001, Bombaci2016_IJMPD-1730004}, which may also take various geometrical structures that are sensitive to the interface effects~\cite{Heiselberg1993_PRL70-1355, Voskresensky2002_PLB541-93, Tatsumi2003_NPA718-359, Voskresensky2003_NPA723-291, Endo2005_NPA749-333, Maruyama2007_PRD76-123015, Yasutake2014_PRC89-065803, Xia2019_PRD99-103017, Maslov2019_PRC100-025802, Xia2020_PRD102-023031}. Additionally, the spontaneous chiral symmetry breaking~\cite{Buball2005_PR407-205} and formation of color superconducting phases in quark matter~\cite{Alford2008_RMP80-1455, Ruester2005_PRD72-034004} could also lead to first-order phase transitions and form various mixed phases~\cite{Buballa2015_PPNP81-39, Voskresensky2023_PPNP130-104030}. In fact, recent Bayesian analysis~\cite{Xie2021_PRC103-035802, Annala2023_NC14-8451, Han2023_SB68-913, Pang2024_PRC109-025807} and binary neutron star merger simulations~\cite{Bauswein2019_PRL122-061102, Huang2022_PRL129-181101, Fujimoto2023_PRL130-91404} have shown great potential on identifying a possible first-order (deconfinement) phase transition in massive compact stars. It is thus interesting to investigate the properties of those mixed phases and fix the corresponding surface tension values, which could be important to the structures and evolutions of compact stars.

In our previous study we have extended the Nambu-Jona-Lasinio (NJL) model to describe baryonic matter, quark matter, and their transitions in a unified manner, where various first-order phase transitions such as quarkyonic, chiral, and deconfinement phase transitions were identified~\cite{Xia2024_PRD110-014022}. In this work, based on the extended NJL model, we fix the meson masses according to the binding energies of finite nuclei, which affects the range of meson exchange interactions. We then investigate the properties and microscopic structures of their mixed phases and estimate the corresponding surface tension. In particular, adopting a similar numerical recipe as in our previous investigations~\cite{Xia2022_PRC105-045803, Xia2022_PRD106-063020, Xia2022_CTP74-095303}, we examine the possible formation of various geometrical structures based on the Thomas-Fermi approximation. The nuclear pasta in the crust region of compact stars and the mixed phases corresponding to the first-order chiral and deconfinement phase transitions in the core regions of compact stars are fixed, while the correspond surface tensions were estimated adopting a thin-wall approximation~\cite{Avancini2010_PRC82-055807}.

The paper is organized as follows. In Sec.~\ref{sec:the} we present our theoretical framework, including the Lagrangian density of the extended NJL model~\cite{Xia2024_PRD110-014022}, the formalism of the model under the Thomas-Fermi approximation, as well as the formulae to fix the microscopic structures of the mixed phases. The obtained properties and microscopic structures of finite nuclei, nuclear pasta, and mixed phases of various phase transitions are presented in Sec.~\ref{sec:res}, while the corresponding surface tension value are estimated.  We draw our conclusion in Sec.~\ref{sec:con}.

\section{\label{sec:the}Theoretical framework}
\subsection{\label{sec:the_Lagrangian} Lagrangian density}
The Lagrangian density of the extended NJL model~\cite{Xia2024_PRD110-014022} is fixed by
\begin{eqnarray}
\mathcal{L} &=& \sum_{i} \bar{\psi}_i \left(  i \gamma_\mu D^\mu_i - M_i \right)\psi_{i} - \frac{1}{4} A_{\mu\nu}A^{\mu\nu} \label{eq:Lagrangian} \\
 &&\mbox{} - \frac{1}{4} \omega_{\mu\nu}\omega^{\mu\nu}  + \frac{1}{2}m_\omega^2 \omega^2 - \frac{1}{4} \vec{\rho}_{\mu\nu}\cdot\vec{\rho}^{\mu\nu}  + \frac{1}{2}m_\rho^2 \rho^2   \nonumber \\
 &&\mbox{}  + \frac{1}{2}\sum_{i=u,d,s} \left[
\partial_\mu \sigma_i \partial^\mu \sigma_i - m_\sigma^2 \sigma_i^2 \right] +4K
 \bar{n}_{u}^{s}\bar{n}_{d}^{s}\bar{n}_{s}^{s}, \nonumber
\end{eqnarray}
where $\psi_{i}$ represents the Dirac spinor for different fermions $i$ (baryons, quarks and leptons). The field tensors $\omega_{\mu\nu}$, $\vec{\rho}_{\mu\nu}$, and $A_{\mu\nu}$ are
\begin{eqnarray}
\omega_{\mu\nu} &=& \partial_\mu \omega_\nu - \partial_\nu \omega_\mu,  \\
\vec{\rho}_{\mu\nu} &=& \partial_\mu \vec{\rho}_\nu - \partial_\nu \vec{\rho}_\mu,  \\
A_{\mu\nu} &=& \partial_\mu A_\nu - \partial_\nu A_\mu.
\end{eqnarray}
The covariant derivative in Eq.~(\ref{eq:Lagrangian}) reads
\begin{equation}
i D^\mu_i = i \partial^\mu -  f_i g_{\omega} \sum_{q=u,d,s} N^q_i \omega^\mu  - f_i g_{\rho} \vec{\tau}_i\cdot\vec{\rho}^\mu - e q_i A^\mu,
\end{equation}
where $N^q_i$ is the number of valence quarks $q$ in particle $i$, $\vec{\tau}_i$ the isospin, and $q_i$ the charge number of particle $i$ with $q_{n}=q_{\Lambda}=0$, $q_{p}=-q_{e}=-q_{\mu}=1$, $q_{u}=2/3$, and $q_{d}=q_{s}=-1/3$. The parameter $f_i$ modulates the coupling strengths between particle $i$ and vector mesons, where we fix the coupling constants $g_{\omega}$ and $g_{\rho}$ according to nuclear matter properties by taking $f_p=f_n =1$. The baryons are considered as clusters with their effective masses given by
\begin{equation}
 M_i = \sum_{q=u,d,s} N^q_i \left[m_{q0} + \alpha_S(M_{q}-m_{q0})\right]  + B n_\mathrm{b}^Q \label{eq:Bmass}  
\end{equation}
with the quark mass
\begin{equation}
 M_{q} = m_{q0} - g_{\sigma} \sigma_q + 2 K \frac{\bar{n}_{u}^{s}\bar{n}_{d}^{s}\bar{n}_{s}^{s}}{\bar{n}_q^{s}} \label{eq:qmass}
\end{equation}
and the baryon number density of quarks $n_\mathrm{b}^Q=(n_u+n_d+n_s)/3$. Note that the last term in Eq.~(\ref{eq:Bmass}) accounts for the Pauli blocking effects in resemblance to that of $\alpha$ clustering inside nuclear medium~\cite{Roepke2014_PRC90-034304, Xu2016_PRC93-011306}. The effective quark scalar density is determined by
\begin{equation}
  \bar{n}_q^s= n_{q}^{s}+\alpha_S \sum_{i=p,n,\Lambda} N^q_i  n_{i}^{s}, \label{eq:nqs}
\end{equation}
which contains the contributions of quasi-free quarks and baryons and eventually vanishes at large enough densities. The scalar and vector densities in Eqs.~(\ref{eq:Bmass})-(\ref{eq:nqs}) are fixed by Eqs.~(\ref{eq:2.nv}-\ref{eq:ns}). Note that the masses of leptons remain constant with $M_e=0.511~$MeV and $M_{\mu}=105.66~$MeV~\cite{Olive2014_CPC38-090001}.

To reproduce the baryon masses in vacuum and medium, we have introduced a density dependent structural function $\alpha_S$, i.e.,
\begin{equation}
   \alpha_S = a_S \exp(-n_\mathrm{b}/n_S)+b_S,  \label{eq:alphaS}
\end{equation}
which accounts for the dampened interaction strength as chiral condensates diminish inside baryons~\cite{Bentz2001_NPA696-138, Reinhardt2012_PRD85_074029, Xia2014_CPL31-41101}. The density dependent couplings for vector mesons are adopted as well to reproduce the nuclear saturation properties, i.e.,
\begin{eqnarray}
g_\omega^2 &=& 4G_S m_\omega^2 [a_V\exp(-n_\mathrm{b}/n_V) + b_V], \label{eq:alphaV}\\
g_\rho^2   &=& 4G_S m_\rho^2 [a_{TV}\exp(-n_\mathrm{b}/n_{TV}) + b_{TV}]. \label{eq:alphaTV}
\end{eqnarray}
The corresponding coefficients in Eqs.~(\ref{eq:alphaS})-(\ref{eq:alphaTV}) are indicated in Table.~\ref{table:DDparam}.

In the mean-field approximation (MFA), the boson fields $\sigma_i$, $\omega_\mu$, $\vec{\rho}_\mu$, and $A_\mu$ take mean values and are left with the time components due to time-reversal symmetry, while charge conservation demands that only the $3$rd component (${\rho}_{\mu,3}$ and $\tau_{i,3}$) in the isospin space remains. We then define $\omega\equiv \omega_0$, $\rho\equiv \rho_{0,3}$, and $\tau_i\equiv \tau_{i,3}$ for simplicity. Based on the Lagrangian density in Eq.~(\ref{eq:Lagrangian}), the meson and photon fields in MFA are determined by
\begin{eqnarray}
(-\nabla^2 + m_\sigma^2) \sigma_q &=&  g_{\sigma} n_{q}^{s} + \alpha_S g_{\sigma}\sum_{i=p,n,\Lambda} N^q_i n_{i}^{s}, \label{eq:KG_sigma} \\
(-\nabla^2 + m_\omega^2) \omega &=& g_{\omega} \sum_i f_i \sum_{q=u,d,s} N^q_i n_i, \label{eq:KG_omega}\\
(-\nabla^2 + m_\rho^2) \rho     &=& g_{\rho} \sum_i f_i  \tau_i n_i, \label{eq:KG_rho}\\
                   -\nabla^2 A  &=& e\sum_{i}q_i n_i, \label{eq:KG_photon}
\end{eqnarray}
where the source currents of fermion $i$ are $n_{i} = \langle \bar{\psi}_{i} \gamma^{0} \psi_{i} \rangle$ and $n_{i}^{s} = \langle \bar{\psi}_{i} \psi_{i} \rangle$.

\subsection{\label{sec:the_TFA} Thomas-Fermi approximation}
Adopting the Thomas-Fermi approximation (TFA), the local source currents of fermion $i$ for cold dense matter are obtained with
\begin{eqnarray}
n_{i}&=&\langle \bar{\psi}_{i} \gamma^{0} \psi_{i} \rangle = \frac{d_{i}\nu_{i}^{3}}{6\pi^{2}}, \label{eq:2.nv}\\
n_{i}^{s}&=&\langle \bar{\psi}_{i} \psi_{i} \rangle = \frac{d_i M_{i}^{3}}{4\pi^{2}}\left[x_{i}\sqrt{x_{i}^{2}+1}-\mathrm{arcsh}(x_{i}) \right.  \nonumber \\
&&   \left. - y_{i}\sqrt{y_{i}^{2}+1}+\mathrm{arcsh}(y_{i})\right]. \label{eq:ns}
\end{eqnarray}
Here we have defined $x_{i} \equiv \nu_{i}/M_{i}$ with $\nu_{i}$ being the Fermi momentum, $y_{i} \equiv \lambda/M_{i}$ with $\lambda$ being the 3-momentum cut-off to regularize the vacuum part of quarks ($y_{i}=0$ for baryons and leptons), and $d_{n,p}=d_{e,\mu}=2$ and $d_{u,d,s}=6$ the degeneracy factors.

The total energy of the system is then fixed by
\begin{equation}
E=\int \langle {\cal{T}}_{00} \rangle \mbox{d}^3 r, \label{eq:energy}
\end{equation}
with the energy momentum tensor
\begin{eqnarray}
\langle {\cal{T}}_{00} \rangle
&=& \sum_{i} \frac {d_i{M_i}^4}{16\pi^{2}} \left[x_i(2x_i^2+1)\sqrt{x_i^2+1}-\mathrm{arcsh}(x_i) \right] \label{eq:ener_dens} \\
&&   + \frac{1}{2}\left[(\nabla \omega)^2 + m_\omega^2 \omega^2  + (\nabla \rho)^2 + m_\rho^2 \rho^2 +(\nabla A)^2\right]  \nonumber \\
&&  -\sum_{i=u,d,s} \frac {d_i{M_i}^4}{16\pi^{2}} \left[y_i(2y_i^2+1)\sqrt{y_i^2+1}-\mathrm{arcsh}(y_i) \right] \nonumber \\
&&     -4K \bar{n}_{u}^{s}\bar{n}_{d}^{s}\bar{n}_{s}^{s} - \mathcal{E}_0 + \frac{1}{2}\sum_{i=u,d,s}\left[(\nabla \sigma_i)^2 +m_\sigma^2 \sigma_i^2\right]. \nonumber
\end{eqnarray}
Here a constant $\mathcal{E}_0$ is introduced to ensure $\langle {\cal{T}}_{00} \rangle = 0$ in the vacuum. The optimum density distributions $n_i(\vec{r})$ of fermions in TFA are fixed by minimizing the total energy $E$ at given total particle numbers $N_i=\int n_i \mbox{d}^3 r$, dimension $D$, and WS cell size $R_\mathrm{W}$, which follows the constancy of local chemical potentials  $\mu_i(\vec{r})$ fixed by
\begin{eqnarray}
 \mu_b &=& \sqrt{\nu_b^{2}+M_b^{2}} + f_b(3  g_{\omega} \omega +  g_{\rho} \tau_{b}  \rho)  +\Sigma_b^\mathrm{R}+e q_b  A, \label{eq:chem_b}\\
 \mu_{q} &=&\sqrt{\nu_q^{2}+M_{q}^2}   + g_{\omega} \omega +g_{\rho} \tau_q \rho + \Sigma_{q}^\mathrm{R}+ e q_q  A, \label{eq:chem_q}\\
 \mu_{l} &=&\sqrt{\nu_l^{2}+m_{l}^2} - e  A, \label{eq:chem_l}
\end{eqnarray}
with the ``rearrangement'' terms given by
\begin{eqnarray}
 \Sigma_b^\mathrm{R}&=& \sum_{i} f_i\left( \omega n_{i}
 \sum_{q=u,d,s} N^q_i \frac{\mbox{d} g_{\omega}}{\mbox{d} n_\mathrm{b}} +  \rho \tau_{i} n_{i} \frac{\mbox{d} g_{\rho}}{\mbox{d} n_\mathrm{b}}\right)
  \nonumber \\
 && +\sum_{i=n,p,\Lambda} \left[ \frac{\mbox{d}  \alpha_S}{\mbox{d} n_\mathrm{b}}\sum_{q=u,d,s} N^q_i(M_{q}-m_{q0}) \right] n_{i}^s, \label{eq:Sigma_b} \\
 \Sigma_{q}^\mathrm{R}&=& \frac{1}{3} B \sum_{i=n,p,\Lambda}n_{i}^s + \frac{1}{3}\Sigma_b^\mathrm{R}. \label{eq:Sigma_Q}
\end{eqnarray}

\subsection{\label{sec:the_EOS} Microscopic structures of mixed phases}
Due to first-order phase transitions, various microscopic structures of their mixed phases may emerge at different densities. For example, a liquid-gas mixed phase of nuclear matter may be formed at $n_\mathrm{b}\lesssim 0.08\ \mathrm{fm}^{-3}$, which are typically found in the crusts of neutron stars. Various shapes and lattice structures are expected due to the interplay between the Coulomb and strong interactions~\cite{Baym1971_ApJ170-299, Negele1973_NPA207-298, Ravenhall1983_PRL50-2066, Hashimoto1984_PTP71-320, Williams1985_NPA435-844}. At larger densities, quarkyonic, chiral, and deconfinement phase transitions are expected to take place~\cite{Xia2024_PRD110-014022}, which may become of first-order and could form various geometrical structures for the corresponding mixed phases~\cite{Heiselberg1993_PRL70-1355, Glendenning2001_PR342-393, Voskresensky2002_PLB541-93, Tatsumi2003_NPA718-359, Voskresensky2003_NPA723-291, Endo2005_NPA749-333, Maruyama2007_PRD76-123015, Peng2008_PRC77-065807, Yasutake2014_PRC89-065803, Maslov2019_PRC100-025802, Xia2019_PRD99-103017, Xia2020_PRD102-023031}.

To obtain the microscopic structures of the mixed phases, we solve the Klein-Gordon equations and fix the density distributions iteratively inside a WS cell. Adopting the spherical and cylindrical approximations~\cite{Maruyama2005_PRC72-015802}, the derivatives in the Klein-Gordon equations~(\ref{eq:KG_sigma}-\ref{eq:KG_photon}) can be reduced into one-dimensional, i.e.,
\begin{eqnarray}
 \mathrm{1D:}\ \ \ \  && \nabla^2 \phi(\vec{r}) = \frac{\mbox{d}^2\phi(r)}{\mbox{d}r^2}; \label{eq:dif_1D} \\
 \mathrm{2D:}\ \ \ \  && \nabla^2 \phi(\vec{r}) = \frac{\mbox{d}^2\phi(r)}{\mbox{d}r^2} + \frac{1}{r} \frac{\mbox{d}\phi(r)}{\mbox{d}r}; \label{eq:dif_2D}\\
 \mathrm{3D:}\ \ \ \  && \nabla^2 \phi(\vec{r}) = \frac{\mbox{d}^2\phi(r)}{\mbox{d}r^2} + \frac{2}{r} \frac{\mbox{d}\phi(r)}{\mbox{d}r}; \label{eq:dif_3D}
\end{eqnarray}
which are solved via fast cosine transformation fulfilling the reflective boundary conditions at $r=0$ and $r=R_\mathrm{W}$~\cite{Xia2021_PRC103-055812}. Adopting imaginary time step method~\cite{Levit1984_PLB139-147}, the density distributions of fermions are obtained with Eqs.~(\ref{eq:chem_b}-\ref{eq:chem_l}) fulfilling the $\beta$-stability condition, i.e.,
\begin{equation}
\mu_i= B_i \mu_\mathrm{B} - q_i \mu_e.  \label{eq:weakequi}
\end{equation}
where $\mu_\mathrm{B}$ and $\mu_e$ take constant values with the baryon number $B_p=B_n=B_\Lambda=1$, $B_u=B_d=B_s=1/3$, and $B_e=B_\mu=0$. Note that at each iteration, the total particle numbers satisfy the global charge neutrality condition
\begin{equation}
  \sum_{i} \int q_i n_i(\vec{r}) \mbox{d}^3 r\equiv 0. \label{eq:globalch}
\end{equation}
Different types of microscopic structures can be obtained with Eqs.~(\ref{eq:dif_1D}-\ref{eq:dif_3D}), i.e., droplet, rod, slab, tube, bubble, and uniform. At a given average baryon number density $n_\mathrm{b}$, we then search for the energy minimum among six types of microscopic structures with optimum cell sizes $R_\mathrm{W}$. The droplet size $R_\mathrm{d}$ is determined by
\begin{equation}
 R_\mathrm{d} =
 \left\{\begin{array}{l}
   R_\mathrm{W}\left(\frac{\langle n_\mathrm{I} \rangle^2}{\langle n_\mathrm{I}^2 \rangle}\right)^{1/D},  \text{\ \ \ \ \ \ \  droplet-like}\\
   R_\mathrm{W} \left(1- \frac{\langle n_\mathrm{I} \rangle^2}{\langle n_\mathrm{I}^2 \rangle}\right)^{1/D},  \text{\ \ bubble-like}\\
 \end{array}\right.,  \label{Eq:Rd}
\end{equation}
where $\langle n_\mathrm{I}^2 \rangle = \int [n_\mathrm{I}(\vec{r})]^2 \mbox{d}^3 r/V$ and $\langle n_\mathrm{I} \rangle  = \int n_\mathrm{I}(\vec{r}) \mbox{d}^3 r/V$ with $n_\mathrm{I}=n^\mathrm{Q}_\mathrm{b}$ or $n_p$ for quark droplets or nuclei, respectively. The WS cell volume corresponds to the cell size $R_\mathrm{W}$ via
\begin{equation}
  V =
 \left\{\begin{array}{l}
   \frac{4}{3}\pi R_\mathrm{W}^3,\  D = 3\\
   \pi a R_\mathrm{W}^2 , \  D = 2\\
   a^2 R_\mathrm{W}, \ \  D = 1\\
 \end{array}\right.. \label{Eq:V}
\end{equation}
In order for the volume to be finite for the slabs and rods/tubes at $D = 1$ and 2, a finite cell size $a$ is usually adopted.

\section{\label{sec:res}Results and discussions}

\subsection{\label{sec:nucl}Finite nuclei}

\begin{table}
  \centering
  \caption{\label{table:DDparam} Adopted parameters of the extended NJL model~\cite{Xia2024_PRD110-014022}, where the RKH parameter set is adopted to reproduce various meson properties, i.e., $\lambda = 602.3$ MeV, $m_{u0}= m_{d0} = 5.5$ MeV,  $m_{s0}=  140.7$ MeV, $G_S = g_{\sigma}^2/4m_{\sigma}^2= 1.835/\lambda^2$, $K = 12.36/\lambda^5$~\cite{Rehberg1996_PRC53-410}. The parameters for the density dependent coupling constants in Eqs.~(\ref{eq:alphaS}-\ref{eq:alphaTV}) is adopted to reproduce nuclear saturation properties with $f_p=f_n=1$  and $f_\Lambda =1.0626$. The meson masses are fixed to reproduce the binding energyies of finite nuclei in TFA.}
  \begin{tabular}{l|l|l}
    \hline \hline
  $a_S=0.4413715$     & $n_S=0.16\ \mathrm{fm}^{-3}$    & $b_S=0.4076285$ \\
  $a_V=3.566049$     & $n_V=0.214\ \mathrm{fm}^{-3}$   & $b_V=1.062771$ \\
  $a_{TV}=0.5014459$  & $n_{TV}=0.1\ \mathrm{fm}^{-3}$  & $b_{TV}=0.0117601$ \\     \hline
  $m_{\sigma}=630$ MeV  & $m_{\omega}=10^5$ MeV  & $m_{\rho}=769$ MeV  \\        \hline
  \end{tabular}
\end{table}

\begin{figure}[!ht]
  \centering
  \includegraphics[width=\linewidth]{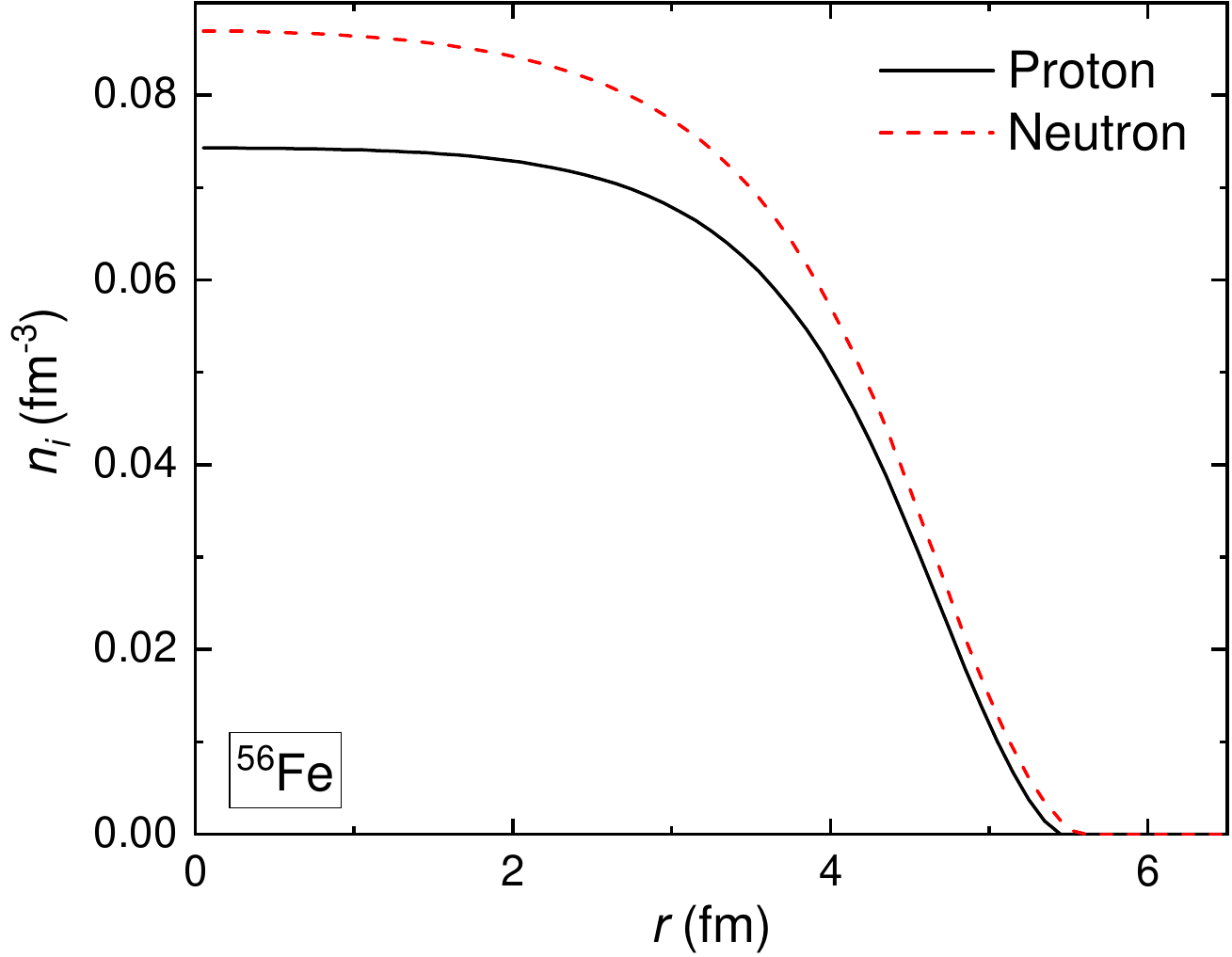}
  \caption{\label{Fig:Fe_Dens}Density profiles of $^{56}$Fe obtained with the extended NJL model in the framework of TFA, where the parameters indicated in Table.~\ref{table:DDparam} are adopted.}
\end{figure}

In our previous study~\cite{Xia2024_PRD110-014022}, the density dependent coupling strengths are fixed by reproducing the saturation properties of nuclear matter as well as the $\Lambda$ potential well depth $U_\Lambda(n_0) = - 30$~MeV in symmetric nuclear matter, while the couplings only appear in combinations with the meson masses as indicated in Eqs.~(\ref{eq:alphaV}) and (\ref{eq:alphaTV}). In this work, we further fix the meson masses $m_{\sigma, \omega, \rho}$ according to the properties of finite nuclei. For a nucleus with fixed proton ($Z$) and neutron ($N$) numbers, in the framework of TFA, we solve the Klein-Gordon equations~(\ref{eq:KG_sigma}-\ref{eq:KG_photon}) and fix the density profiles with Eq.~(\ref{eq:chem_b}) iteratively. Once convergency is reached, the mass of the nucleus is determined by Eq.~(\ref{eq:energy}). As an example, in Fig.~\ref{Fig:Fe_Dens} we illustrate the obtained density profiles of $^{56}$Fe, where the meson masses adopted in this work are indicated in Table.~\ref{table:DDparam}. Note that we have adopted a rather large $\omega$ meson mass $m_{\omega}=10^5$ MeV to prevent density fluctuations caused by the six-point (`t Hooft) interaction in the framework of TFA, where the local density may exceed several times the saturation density and become discontinues if $m_{\omega}$ is not large enough. This situation may be improved if we consider the quantum effects of quarks and baryons by solving the Dirac equations to obtain the wave functions instead of adopting TFA~\cite{Xia2022_PRD106-034016}, so that $m_{\omega}$ can be reduced to more reasonable values.

\begin{figure}[!ht]
  \centering
  \includegraphics[width=\linewidth]{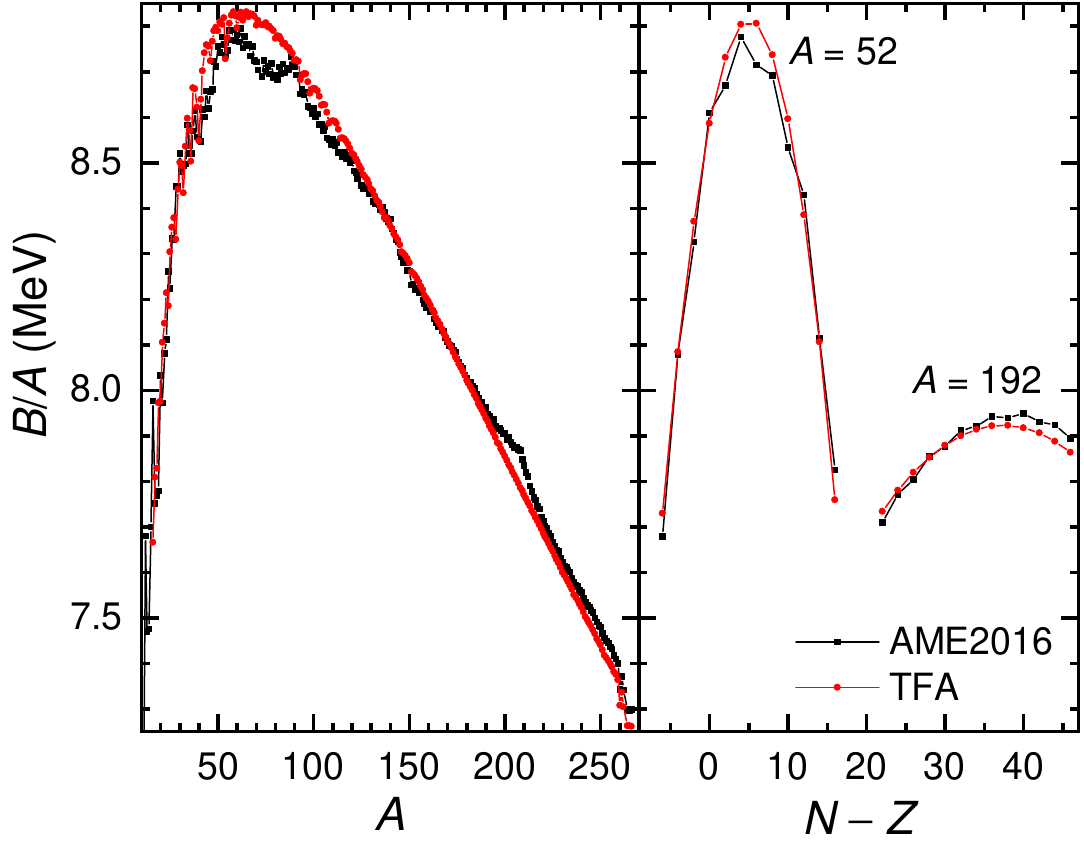}
  \caption{\label{Fig:Bind_Nucl}Binding energy per nucleon of finite nuclei obtained with the extended NJL model in the framework of TFA, where the left panel corresponds to $\beta$-stable nuclei and the right one to nuclei with fixed mass numbers $A$. The results are compared with the experimental data from AME2016~\cite{Audi2017_CPC41-030001, Huang2017_CPC41-030002, Wang2017_CPC41-030003}.}
\end{figure}

Carrying out systematic calculations for $\beta$-stable nuclei and nuclei with fixed mass numbers $A=52$ and 192 within the framework of TFA, their properties can then be obtained. In Fig.~\ref{Fig:Bind_Nucl} the corresponding binding energy per nucleon ($B/A$) are presented and compared with the experimental values from the 2016 Atomic Mass Evaluation (AME2016)~\cite{Audi2017_CPC41-030001, Huang2017_CPC41-030002, Wang2017_CPC41-030003}. Here we take $m_{\sigma}=630$ MeV to reproduce the binding energy of $\beta$-stable nuclei, which is slightly heavier than the typical $\sigma$ meson masses. 
The predicted binding energies for $\beta$-stable nuclei are indicated in the left panel of Fig.~\ref{Fig:Bind_Nucl}, which are generally consistent with the experimental values with slight deviations around the magic numbers. The deviations can be reduced if we further consider the shell effects and nucleon pairing. For the $\rho$ meson mass, we take a typical value $m_{\rho}=769$ MeV, which gives satisfactory description for the binding energies of nuclei at various proton and neutron numbers as indicated in the right panel of Fig.~\ref{Fig:Bind_Nucl}.

\subsection{\label{sec:mix}Mixed phases}
The mixed phases of first-order phase transitions inside compact stars could form various geometrical structures. We thus investigate the corresponding microscopic structures based on the numerical recipe indicated in Sec.~\ref{sec:the_EOS}, where the spherical and cylindrical approximations~\cite{Maruyama2005_PRC72-015802} for the WS cells are adopted.

\subsubsection{\label{sec:lg}Nuclear pasta}
At small densities $n_\mathrm{b}< 0.09\ \mathrm{fm}^{-3}$, nuclear matter exhibits a first-order liquid-gas phase transition in the absence of an early emergence of quaryonic phase. A liquid-gas mixed phase of nuclear matter is then formed, which are typically found in the crusts of neutron stars. We then search for the optimum structures that minimize the energy per baryon at a fixed average baryon number density $n_\mathrm{b}$.

\begin{figure}[!ht]
  \centering
  \includegraphics[width=\linewidth]{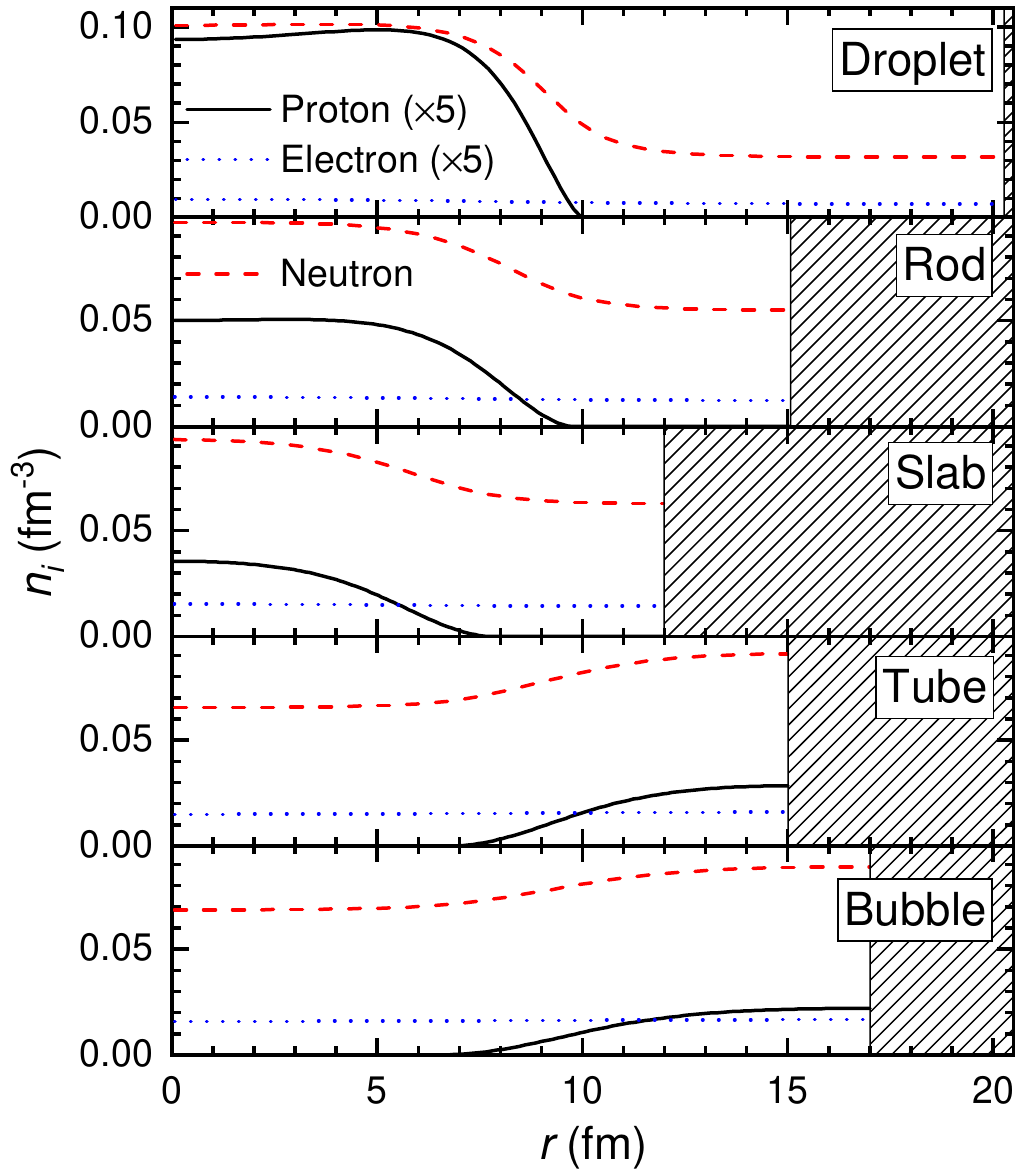}
  \caption{\label{Fig:Pasta_Str} Density profiles of nucleons and electrons in the WS cells for droplet, rod, slab, tube, and bubble phases at $n_\mathrm{b}=0.04$, 0.07, 0.08, 0.084, 0.088 fm$^{-3}$ (from top to bottom), respectively. The boundary of the WS cell at $r = R_\mathrm{W}$ is indicated by the shaded region.}
\end{figure}

In Fig.~\ref{Fig:Pasta_Str} we show the typical density profiles of nucleons and electrons in the WS cells for droplet, rod, slab, tube, and bubble phases at $n_\mathrm{b}=0.04$, 0.07, 0.08, 0.084, 0.088 fm$^{-3}$, which correspond to the optimum configurations at those densities. Due to the requirement of $\beta$-stability and global charge neutrality conditions indicated in Eqs.~(\ref{eq:weakequi}) and (\ref{eq:globalch}), the proton fraction of nuclear pasta is small and neutrons exist throughout the WS cells. Compared with the uniform neutron star matter, the energy per baryon is reduced by up to $\sim$8 MeV/A due to the formation of nonuniform structures.

\begin{figure}[!ht]
  \centering
  \includegraphics[width=\linewidth]{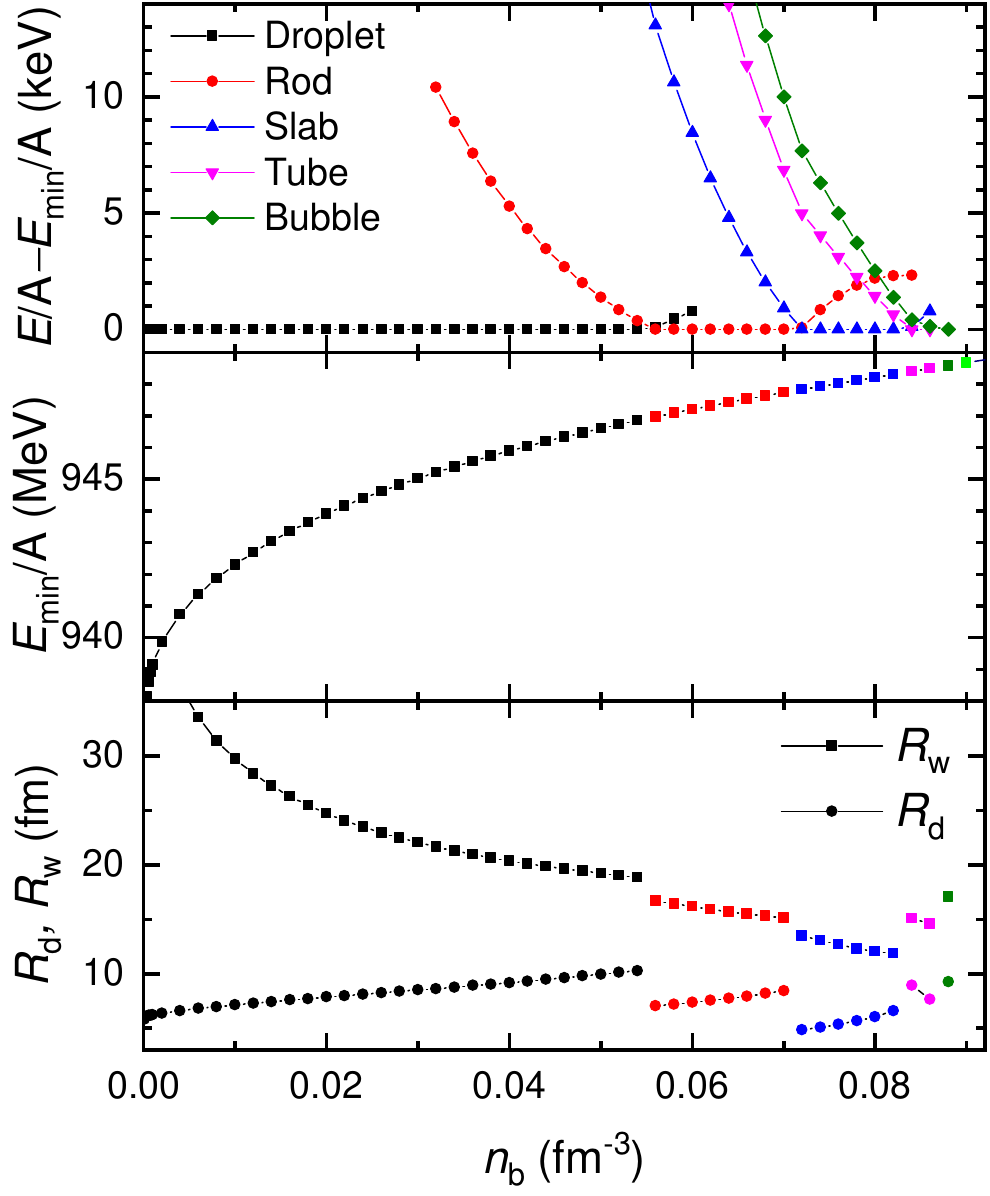}
  \caption{\label{Fig:Mic_pasta} Energy excess per baryon, minimum energy per baryon, WS cell size $R_\mathrm{W}$ and droplet size $R_\mathrm{d}$ for $\beta$-stable nuclear matter inside compact stars.}
\end{figure}

Carrying out a more extensive calculation, we can then obtain the microscopic structures of nuclear pasta for $\beta$-stable nuclear matter throughout the density range at $n_\mathrm{b}\lesssim 0.09\ \mathrm{fm}^{-3}$. In Fig.~\ref{Fig:Mic_pasta} the corresponding energy excess per baryon with respect to the minimum energy per baryon for various types of nuclear pastas are presented, while the optimum WS cell size $R_\mathrm{W}$ and droplet size $R_\mathrm{d}$ are indicated as well. The obtained results generally agrees with our previous investigations on nuclear pasta using RMF models~\cite{Maruyama2005_PRC72-015802, Xia2022_PRC105-045803, Xia2022_CTP74-095303, Xia2022_PRD106-063020}, where the droplet, rod, slab, tube and bubble phases appear sequentially as density increases. The droplet size $R_\mathrm{d}$ increases with density for droplet-like structures, while $R_\mathrm{d}$ decreases with density for the bubble phase. Meanwhile, the WS cell size $R_\mathrm{W}$ decreases with density, suggesting that the volume fraction of the liquid phase is increasing with density, and finally at $n_\mathrm{b}= 0.09\ \mathrm{fm}^{-3}$ neutron star matter becomes isotropic, i.e., the core-crust transition takes place.

\subsubsection{\label{sec:chiral}Mixed phase of chiral phase transition}
At slightly larger densities with $n_\mathrm{b}\approx 2$-4$n_0$, according to our previous estimations, there exist first-order chiral phase transitions with the masses of $u$ and $d$ quarks restored to their current values. We thus examine the structures and properties of the corresponding mixed phases. It is found that the geometrical structures for the mixed phases are unstable, so that the Maxwell construction scenarios are effectively restored, corresponding to the bulk separation of the two phases.

\begin{figure}[!ht]
  \centering
  \includegraphics[width=\linewidth]{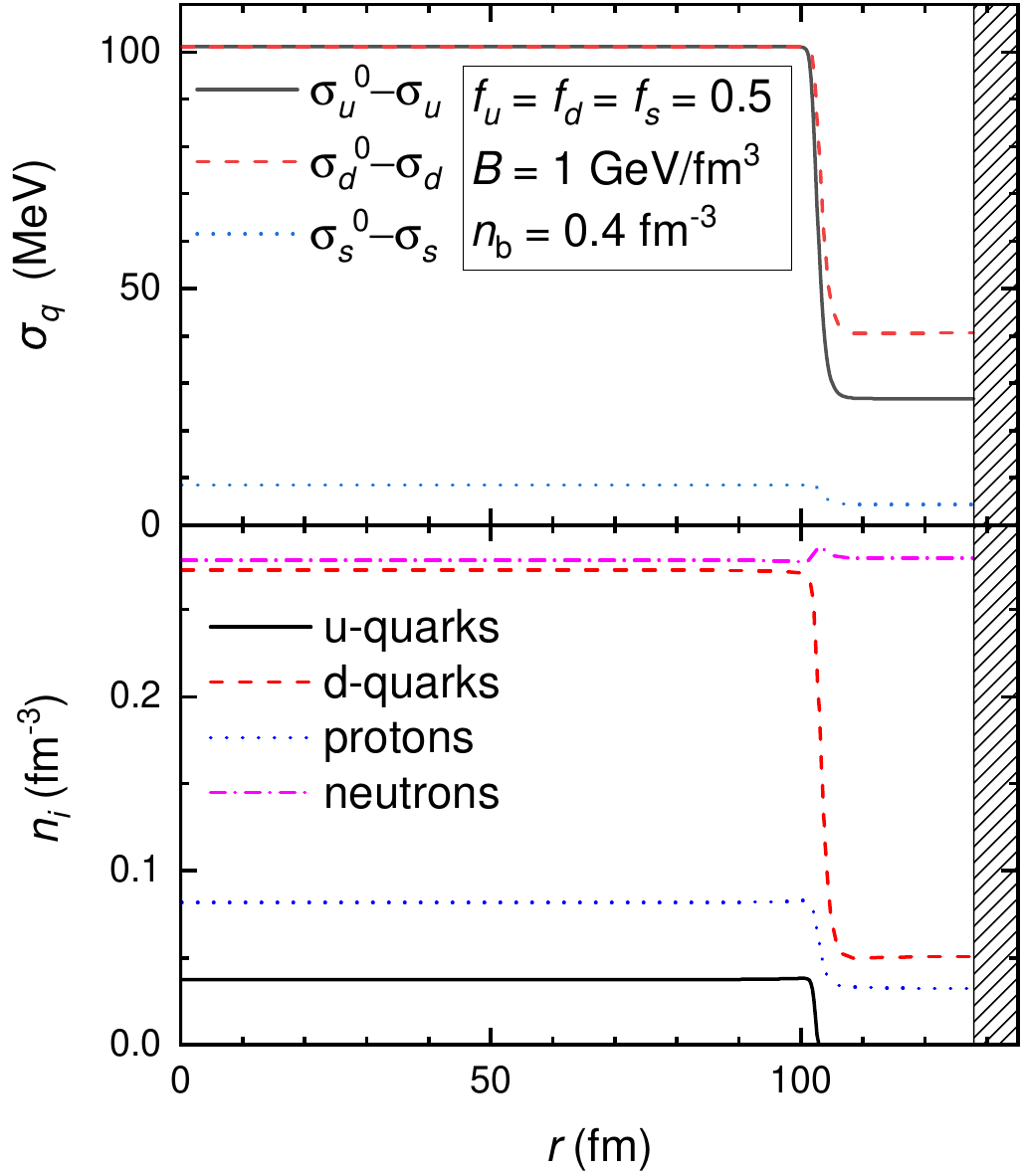}
  \caption{\label{Fig:Dens_f05B1n04} Density profiles of Fermions (lower panel) and the corresponding $\sigma$ meson fields (upper panel) in a WS cell for the droplet phase at $n_\mathrm{b}=0.4$ fm$^{-3}$, where the expectation values of $\sigma$ fields in vacuum are $\sigma_u^0=\sigma_d^0=101.123$ MeV and $\sigma_s^0=122.173$ MeV.}
\end{figure}

As an example, in Fig.~\ref{Fig:Dens_f05B1n04} we present the density profiles of Fermions and the corresponding $\sigma$ meson fields in a WS cell for the droplet phase at $n_\mathrm{b}=0.4$ fm$^{-3}$ in $\beta$-equilibrium, where the parameter set $B = 1$ GeV/fm$^3$ and $f_{u,d,s}=0.5$ is adopted. Evidently, the droplet phase indicated in Fig.~\ref{Fig:Dens_f05B1n04} is comprised of two phases with ($\mathrm{\uppercase\expandafter{\romannumeral2}}$) and without ($\mathrm{\uppercase\expandafter{\romannumeral1}}$) the partial restorations of chiral symmetry for $u$ and $d$ quarks. Note that in both phases strangeness does not emerge due to the large expectation values of $\sigma_s$ arise from the chiral condensate of $s$ quarks, causing $s$ quarks and $\Lambda$ hyperons become too massive to exist in the system. The density for quarks becomes larger with the partial restorations of chiral condensates and a reduction of $\sigma_{u,d}$ ($\rightarrow0$) in phase $\mathrm{\uppercase\expandafter{\romannumeral2}}$, while in phase $\mathrm{\uppercase\expandafter{\romannumeral1}}$ the density for neutrons is larger. At the boundary where two phases contact, the densities and meson fields change smoothly from one phase to the other. The varying meson fields extend approximately 8 fm in and out of the boundary between the two phases, while the Coulomb potential varies on a larger scale with the corresponding Debye screening lengths $\lambda_D^\mathrm{\uppercase\expandafter{\romannumeral1}}\approx \lambda_D^\mathrm{\uppercase\expandafter{\romannumeral2}}\approx 5.5$ fm.

\begin{table*}
  \centering
  \caption{\label{table:MP_prop} Bulk properties for the two phases ($\mathrm{\uppercase\expandafter{\romannumeral1}}$/$\mathrm{\uppercase\expandafter{\romannumeral2}}$) arise from various first-order phase transitions, where the baryon number density $n_\mathrm{b}$, charge fraction of strongly interacting matter $f_z=(3n_p+2n_u-n_d-n_s)/3n_\mathrm{b}$, quark fraction $f_q=(n_u+n_d+n_s)/3n_\mathrm{b}$, and electron number density $n_e$ of each phase are indicated. The surface tensions $\sigma$ are presented as well.}
\begin{tabular}{c c|c c c c|c c c c|c}
  \hline
$B$& $f_{u,d,s}$ & $n_\mathrm{b}^\mathrm{\uppercase\expandafter{\romannumeral1}}$ & $f_z^\mathrm{\uppercase\expandafter{\romannumeral1}}$ & $f_q^\mathrm{\uppercase\expandafter{\romannumeral1}}$ & $n_e^\mathrm{\uppercase\expandafter{\romannumeral1}}$ & $n_\mathrm{b}^\mathrm{\uppercase\expandafter{\romannumeral2}}$ & $f_z^\mathrm{\uppercase\expandafter{\romannumeral2}}$ & $f_q^\mathrm{\uppercase\expandafter{\romannumeral2}}$  &   $n_e^\mathrm{\uppercase\expandafter{\romannumeral2}}$ & $\sigma$      \\
GeV/fm$^3$& & fm$^{-3}$ & & & fm$^{-3}$ & fm$^{-3}$ & &  & fm$^{-3}$ &  MeV/fm$^2$      \\ \hline
 1 &  0.5  &  0.326 &  0.047   & 0.050   & 0.0119 & 0.462   &  0.034  & 0.223 & 0.0121  &    13.4      \\
 1 &  0.5  &  1.26	&  0.0455  & 0.189   & 0.0354 & 1.87    & 2.26$\times10^{-5}$ &  1   &  4.22$\times10^{-5}$ & 29.8        \\
 1 &  0.7  &  0.457	&  0.0910  & 0.00455 & 0.0269  &  0.541 & 0.0586   &  0.130         &  0.0215   &    11.3      \\
 1 &  1    &  0.644 & 0.111    &  0     & 0.0432 &  0.682	&  0.103   & 0.0337         &    0.0425   &    3.43      \\
 0 &  1    &  0.647	&  0.0815  & 0.0583 & 0.0330 &  0.668 & 0.0804   &  0.0645        &    0.0336    &    1.93      \\
  \hline
\end{tabular}
\end{table*}

Based on Fig.~\ref{Fig:Dens_f05B1n04}, we can then extract the corresponding bulk properties of the two phases, which is presented in Table~\ref{table:MP_prop} with phase $\mathrm{\uppercase\expandafter{\romannumeral1}}$/$\mathrm{\uppercase\expandafter{\romannumeral2}}$ being the lower/higher density phase. The properties of the two phases predicted adopting other parameter sets are indicated in Table~\ref{table:MP_prop} as well. Evidently, there exist quasi-free quarks in almost all phases with continues quarkyonic phase transitions take place at lower densities. Nevertheless, if we introduce more repulsive interactions, the quasi-free quarks do not emerge at the lower density phase ($\mathrm{\uppercase\expandafter{\romannumeral1}}$). In particular, if we adopt the parameter set $B = 1$ GeV/fm$^3$ and $f_{u,d,s}=1$, the quarkyonic phase transition and chiral phase transition are entangled and become of first-order, which takes place at approximately 4$n_0$.

In order for the mixed phases to develop geometrical structures, the energy reduction due to the relocation of charged particles should be larger than that of the surface energy. By applying linearization to the charge densities, an analytical formula for the energy reduction per unit surface area can then be derived~\cite{Voskresensky2003_NPA723-291}, i.e.,
\begin{equation}
\sigma_\mathrm{c} = \frac{\left( \mu_{e}^\mathrm{\uppercase\expandafter{\romannumeral1}} - \mu_{e}^\mathrm{\uppercase\expandafter{\romannumeral2}} \right)^2}{8 \pi \alpha \left(\lambda_D^\mathrm{\uppercase\expandafter{\romannumeral1}} + \lambda_D^\mathrm{\uppercase\expandafter{\romannumeral2}}\right)},
\label{eq:sigma_c}
\end{equation}
where $\mu_{e}^{\mathrm{\uppercase\expandafter{\romannumeral1}}, \mathrm{\uppercase\expandafter{\romannumeral2}}}=\sqrt {\left(3 \pi^2 n_e^{\mathrm{\uppercase\expandafter{\romannumeral1}}, \mathrm{\uppercase\expandafter{\romannumeral2}}} \right) ^{2/3}+m_e^2}$ is the local electron chemical potential in the absence of the Coulomb potential with $A=0$ and $\lambda_D^{\mathrm{\uppercase\expandafter{\romannumeral1}}, \mathrm{\uppercase\expandafter{\romannumeral2}}} \equiv \left( -4\pi \alpha \mbox{d}n_\mathrm{ch}^{\mathrm{\uppercase\expandafter{\romannumeral1}}, \mathrm{\uppercase\expandafter{\romannumeral2}}}/\mbox{d}\mu_e\right)^{-1/2}$ the Debye screening lengths of the two phases fulfilling both the $\beta$-stability condition and local charge neutrality condition. The obtained value of $\sigma_\mathrm{c}$ can then be considered as a critical surface tension where geometrical structures emerge only when the surface tension $\sigma<\sigma_\mathrm{c}$. Evidently, as indicated in Table~\ref{table:MP_prop}, the number densities of electrons before and after the chiral phase transitions take similar values with insignificant differences. In such cases, the differences between $\mu_{e}^\mathrm{\uppercase\expandafter{\romannumeral1}}$ and $\mu_{e}^\mathrm{\uppercase\expandafter{\romannumeral2}}$ are tiny, leading to rather small energy reduction according to Eq.~(\ref{eq:sigma_c}), e.g., $\sigma_\mathrm{c} = 0.8$ keV/fm$^2$ if the parameter set $B = 1$ GeV/fm$^3$ and $f_{u,d,s}=0.5$ is adopted.

The surface energy $E_\mathrm{surf}$ of the system can be derived from the grand-canonical potential~\cite{Avancini2010_PRC82-055807}, which takes the form
\begin{equation}
  E_\mathrm{surf} = \int \left[ \sum_{i=u,d,s}\left(\nabla \sigma_i\right)^2 - \left(\nabla \omega \right)^2 - \left(\nabla \rho\right)^2 \right] \mbox{d}^3 r. \label{Eq:Esurf}
\end{equation}
Assuming slab structures for the mixed phases with dimension $D = 1$, we can then estimate the surface tensions $\sigma$ of the interface between the two phases~\cite{Avancini2010_PRC82-055807}, i.e.,
\begin{equation}
  \sigma = \int_{-\infty}^{+\infty}\left[ \sum_{i=u,d,s}\left(\frac{\mbox{d} \sigma_i}{\mbox{d} r}\right)^2 - \left(\frac{\mbox{d} \omega}{\mbox{d} r}\right)^2 - \left(\frac{\mbox{d} \rho}{\mbox{d} r}\right)^2 \right] \mbox{d} r. \label{Eq:sgma}
\end{equation}
The obtained values are then presented in Table~\ref{table:MP_prop}. Evidently, the surface tensions for chiral phase transitions are much larger than the critical surface tension $\sigma_\mathrm{c}$, suggesting that the geometrical structures can not be formed for the corresponding mixed phases.

\begin{figure}[!ht]
  \centering
  \includegraphics[width=\linewidth]{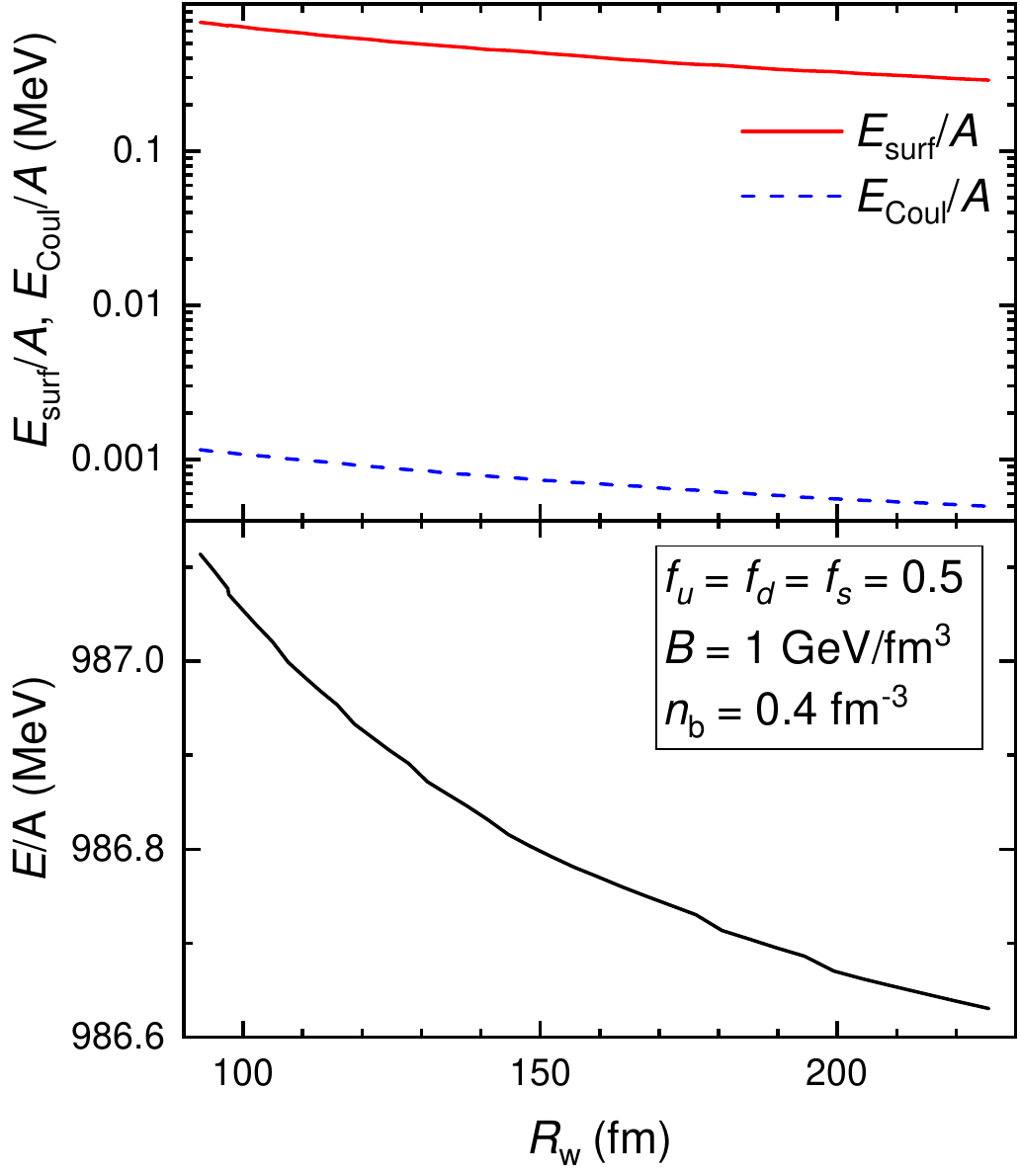}
  \caption{\label{Fig:EpA_f05B1n04} Surface $E_\mathrm{surf}$, Coulomb $E_\mathrm{Coul}$, and total $E$ energies per baryon for the droplet phase corresponding to Fig.~\ref{Fig:Dens_f05B1n04} as functions of WS cell size $R_\mathrm{W}$, which arise from chiral phase transition adopting the parameter set $B = 1$ GeV/fm$^3$ and $f_{u,d,s}=0.5$ at baryon number density $n_\mathrm{b} = 0.4$ fm$^{-3}$.}
\end{figure}

To better illustrate this, in Fig.~\ref{Fig:EpA_f05B1n04} we further examine the variations of surface ($E_\mathrm{surf}$), Coulomb ($E_\mathrm{Coul}$), and total ($E$) energies per baryon for the droplet phase corresponding to Fig.~\ref{Fig:Dens_f05B1n04} as functions of WS cell size $R_\mathrm{W}$, where the Coulomb energy $E_\mathrm{C}$ is obtained with
\begin{equation}
  E_\mathrm{Coul} =  \frac{1}{2}\int (\nabla A)^2\mbox{d}^3 r. \label{Eq:Ecoul}
\end{equation}
It can be seen clearly that the total energy per baryon $E/A$ decreases with $R_\mathrm{W}$ and there does not exist a minimum point for the droplet phase to exist stably. This can also be identified if we examine the surface and Coulomb energies $E_\mathrm{surf}$ and  $E_\mathrm{coul}$. In fact, previous investigations have shown that the inhomogeneous structures are realized by a balance between the surface and Coulomb energies~\cite{Maruyama2005_PRC72-015802, Xia2020_PRD102-023031}, i.e., $E_\mathrm{surf} = 2 E_\mathrm{Coul}$. Evidently, this condition can not be satisfied since $E_\mathrm{surf} \gg E_\mathrm{Coul}$, which is mainly attributed to the small differences between $\mu_{e}^\mathrm{\uppercase\expandafter{\romannumeral1}}$ and $\mu_{e}^\mathrm{\uppercase\expandafter{\romannumeral2}}$.

\begin{figure}[!ht]
  \centering
  \includegraphics[width=\linewidth]{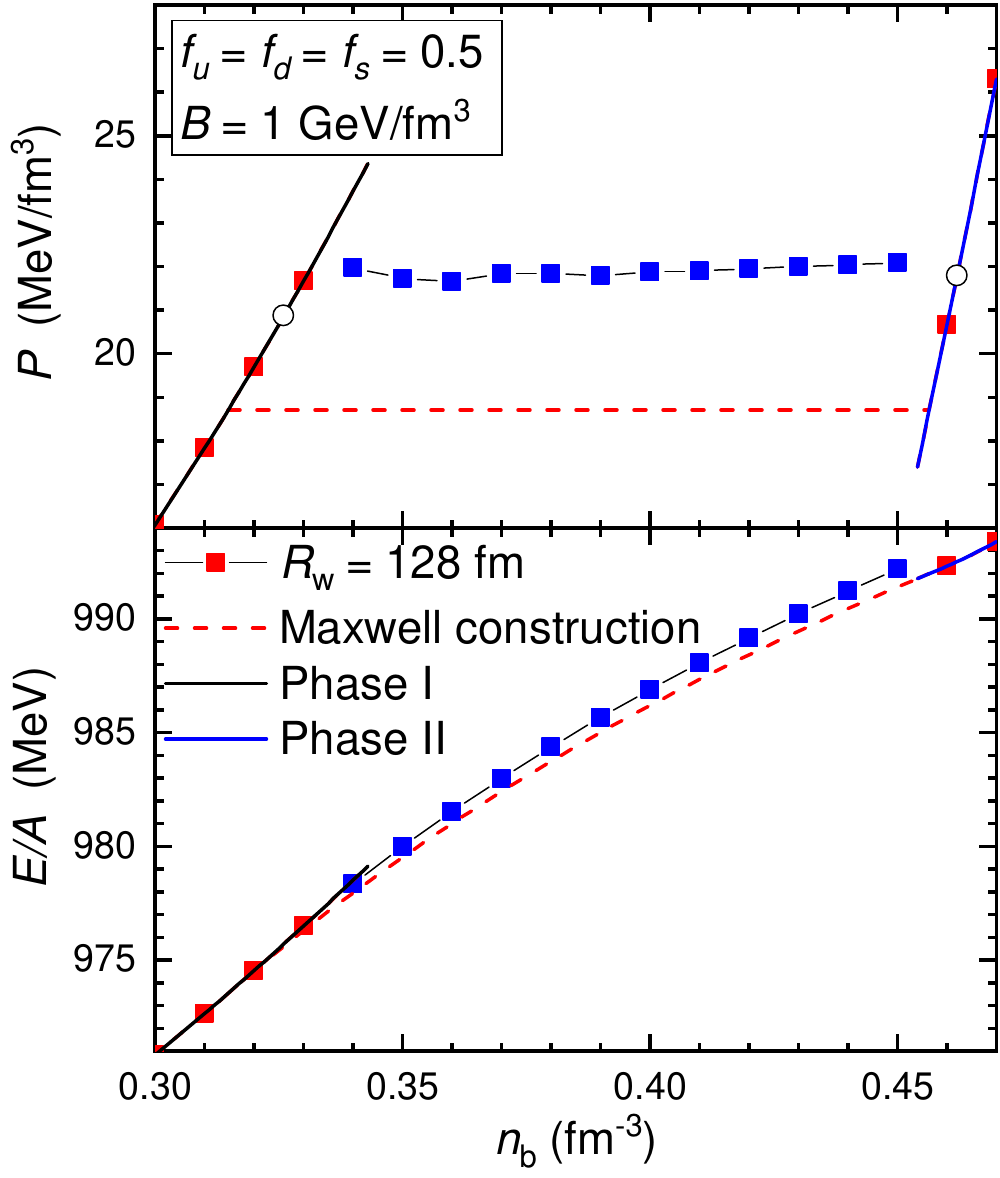}
  \caption{\label{Fig:EOS_f05B1} Pressure $P$ and energy per baryon $E/A$ of the mixed phases for chiral phase transition, where the blue squares indicate the droplet phase obtained at fixed WS cell radius $R_\mathrm{W}=128$ fm. The dashed curve is fixed by employing Maxwell construction for the mixed phase, while the open circles correspond to the bulk properties of phases $\mathrm{\uppercase\expandafter{\romannumeral1}}$ and $\mathrm{\uppercase\expandafter{\romannumeral2}}$ indicated in Table~\ref{table:MP_prop}.}
\end{figure}

Finally, as an example, in Fig.~\ref{Fig:EOS_f05B1} we present the pressure $P$ and energy per baryon $E/A$ of the mixed phases, where the droplet structure indicated in Fig.~\ref{Fig:Dens_f05B1n04} is adopted with a fixed WS cell radius $R_\mathrm{W}=128$ fm for practical calculations. The equation of state (EOS) fixed by employing Maxwell construction is indicated with dashed curves as well. It is found that the variation of pressure for the droplet phase is insignificant as we increase the density, which is similar to that of the Maxwell construction. Nevertheless, we note that the obtained pressure is larger than that of the Maxwell construction. This is mainly attributed to the increased densities of phases $\mathrm{\uppercase\expandafter{\romannumeral1}}$ and $\mathrm{\uppercase\expandafter{\romannumeral2}}$ due to finite-size effects, where the densities $n_\mathrm{b}^\mathrm{\uppercase\expandafter{\romannumeral1}, \uppercase\expandafter{\romannumeral2}}$  of the two phases (0.326 and 0.462 fm$^{-3}$) indicated by the open circles are larger than that of the Maxwell construction (0.315 and 0.456 fm$^{-3}$), i.e., a slightly delayed onset density for chiral phase transitions. Similar situations are observed for other first-order chiral phase transitions adopting different parameter sets, e.g., those indicated in Table~\ref{table:MP_prop}. This can be improved if we adopt larger WS cell radius $R_\mathrm{W}$, e.g., as indicated in Fig.~\ref{Fig:EpA_f05B1n04}, where the energy per baryon, densities, and pressure are expected to be reduced. The stability of the geometric structures can also be identified by examine the energy per baryon, where in the bottom panel of Fig.~\ref{Fig:EOS_f05B1} it is clearly shown that the droplet phase is unstable with respect to the mixed phases fixed by Maxwell construction.

\subsubsection{\label{sec:deconf}Mixed phase of deconfinement phase transition}

For the first-order deconfinement phase transitions, in this work we consider the scenario predicted by adopting the parameter set $B = 1$ GeV/fm$^3$ and $f_{u,d,s}=0.5$, which is entangled with the chiral phase transition for $s$ quarks with the quark condensate $\langle \bar{s}s \rangle\rightarrow 0$. The corresponding bulk properties and surface tension are then presented in Table.~\ref{table:MP_prop}. According to Eq.~(\ref{eq:sigma_c}), the critical surface tension is $\sigma_\mathrm{c} \approx 10^2$ MeV/fm$^2$, which is larger than the surface tension value $\sigma \approx 30$ MeV/fm$^2$ for the interface, indicating the existence of inhomogenous mixed phases for the deconfinement phase transition. We thus carry out more extensive calculations to investigate the microscopic structures for the mixed phases of deconfinement phase transition.

\begin{figure}[!ht]
  \centering
  \includegraphics[width=\linewidth]{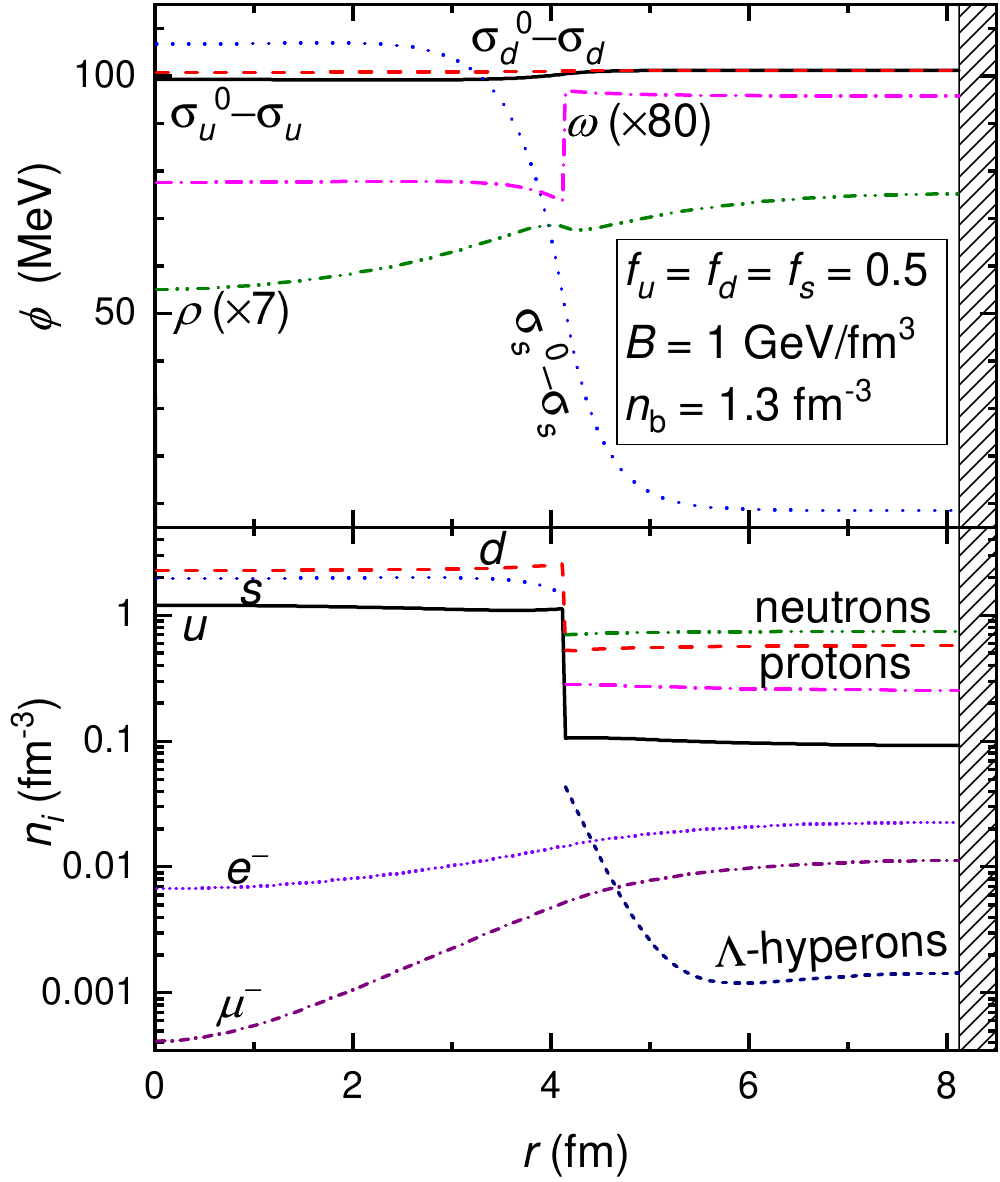}
  \caption{\label{Fig:Dconf_f05B1n13} Density profiles of fermions (lower panel) and the corresponding meson fields (upper panel) in a WS cell for the droplet phase at $n_\mathrm{b}=1.3$ fm$^{-3}$.}
\end{figure}

As an example, in Fig.~\ref{Fig:Dconf_f05B1n13} we present the density profiles of Fermions and the corresponding meson fields in a WS cell for the droplet phase in $\beta$-equilibrium, which is fixed at $n_\mathrm{b}=1.3$ fm$^{-3}$ adopting the parameter set $B = 1$ GeV/fm$^3$ and $f_{u,d,s}=0.5$. Both the confined ($\mathrm{\uppercase\expandafter{\romannumeral1}}$) and deconfined ($\mathrm{\uppercase\expandafter{\romannumeral2}}$) phases can be easily identified. In phase $\mathrm{\uppercase\expandafter{\romannumeral1}}$, there exist various baryons ($p$, $n$, $\Lambda$) and quasi-free quarks ($u$ and $d$), where the chiral condensate $\langle \bar{s}s \rangle$ is still sizable. For the deconfined phase $\mathrm{\uppercase\expandafter{\romannumeral2}}$, strange quark matter (SQM) made of $u$, $d$, $s$ quarks emerge and baryons cease to exist, while the chiral condensate $\langle \bar{s}s \rangle$ vanishes with  $\sigma_s \rightarrow 0$.  Due to the Pauli repulsion adopted in Eq.~(\ref{eq:Bmass}), a sharp interface with sudden changes of densities and $\omega$ field (due to the large $m_\omega$) is identified at the boundary of the two phases, while the other meson fields vary smoothly from one phase to the other. In particular, it is interesting to note that the reduction of $\sigma_s$ in SQM extends beyond phase $\mathrm{\uppercase\expandafter{\romannumeral2}}$ and leads to the attractive interaction between SQM and $\Lambda$-hyperons, which may provide possible signatures for the formation of strangelets~\cite{Chen2024_PRD109-054031}.

\begin{figure}
  \centering
  \includegraphics[width=\linewidth]{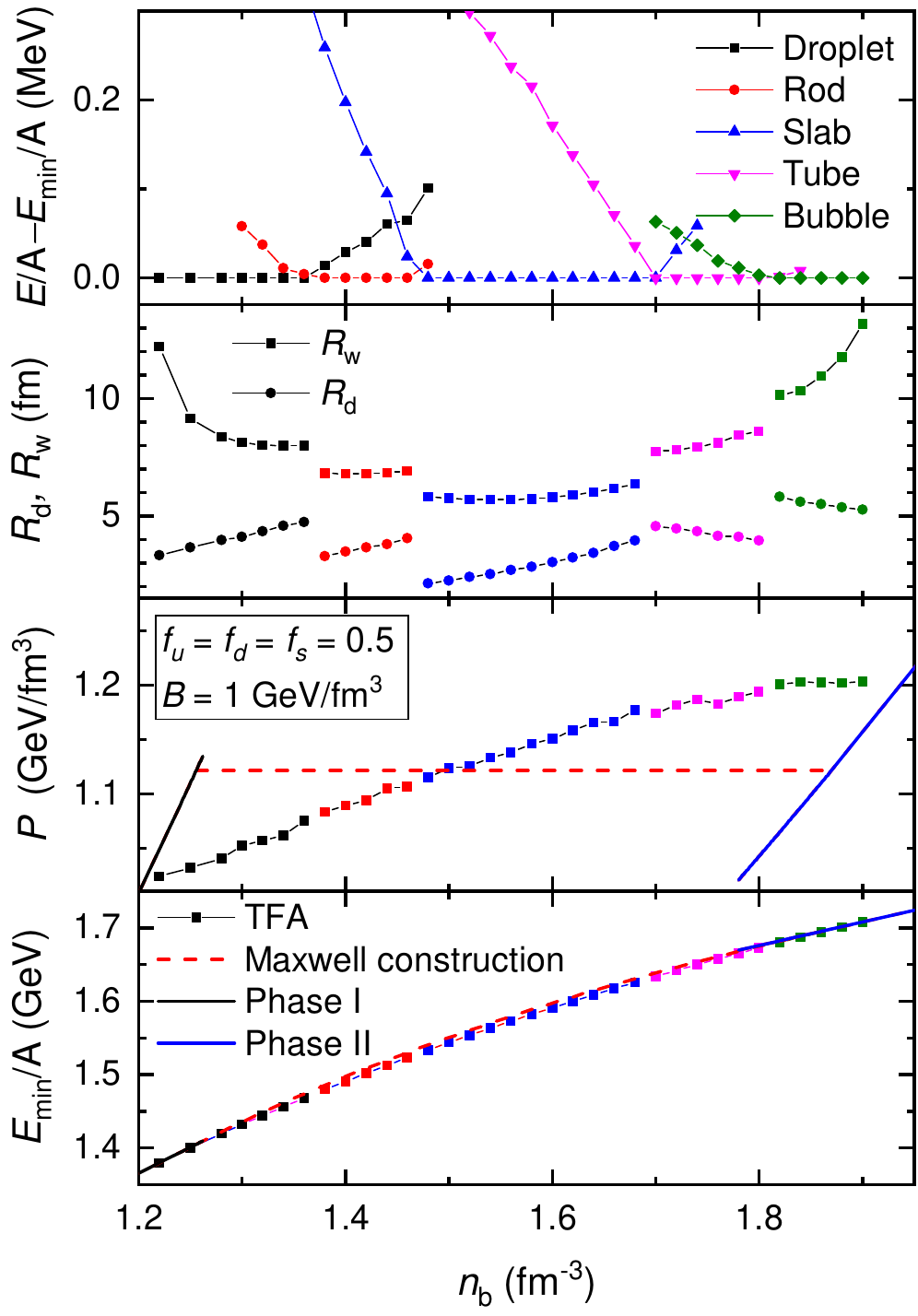}
  \caption{\label{Fig:EOS_Df05B1} Energy excess per baryon, WS cell size $R_\mathrm{W}$ and droplet size $R_\mathrm{d}$, pressure $P$ and minimum energy per baryon $E_\mathrm{min}/A$ of the mixed phases for deconfinement phase transition inside compact stars, while the dashed curve is fixed by employing Maxwell construction for the mixed phase.}
\end{figure}

Finally, in Fig.~\ref{Fig:EOS_Df05B1} we present the energy excess per baryon, WS cell size $R_\mathrm{W}$, droplet size $R_\mathrm{d}$, pressure $P$ and minimum energy per baryon $E_\mathrm{min}/A$ for the most stable configurations of the mixed phases throughout the density region of the deconfinement phase transition. The droplet, rod, slab, tube, and bubble phases emerge sequentially as density increases. The obtained WS cell sizes $R_\mathrm{W}$ are generally small ($\lesssim 10$ fm) with even smaller droplet sizes $R_\mathrm{d}$ as indicated in Fig.~\ref{Fig:Dconf_f05B1n13}. Meanwhile, it is found that $R_\mathrm{d}$ and $R_\mathrm{W}$ decrease as the transitions from droplet to rod and then to slab phases take place, while finally they increase as the tube and bubble phases emerge. Evidently, mixed phases forming geometrical structures are more stable than those obtained with Maxwell construction indicated by the dashed curve, where the corresponding energy per baryon $E_\mathrm{min}/A$ is smaller. The pressure $P$ also deviates from that of the Maxwell construction, which increases with density in contrast to the chiral phase transitions indicated in Fig.~\ref{Fig:EOS_f05B1} with $P$ remains almost constant. It is worth mentioning that in our current work the transitions between different configurations are of first-order and there may exist various exotic configurations~\cite{Magierski2002_PRC65-045804, Watanabe2003_PRC68-035806, Newton2009_PRC79-055801, Nakazato2009_PRL103-132501, Okamoto2012_PLB713-284, Schuetrumpf2013_PRC87-055805, Schneider2014_PRC90-055805, Schuetrumpf2015_PRC91-025801, Sagert2016_PRC93-055801, Berry2016_PRC94-055801, Fattoyev2017_PRC95-055804, Schuetrumpf2019_PRC100-045806, Kashiwaba2020_PRC101-045804}. In such cases, we need to carry out more extensive calculations in a three-dimensional geometry with large box sizes~\cite{Okamoto2012_PLB713-284, Okamoto2013_PRC88-025801, Xia2021_PRC103-055812, Xia2022_PRD106-063020}, which should be done in our future study.

\section{\label{sec:con}Conclusion}

Based on an extended NJL model that treats baryons as clusters of quarks~\cite{Xia2024_PRD110-014022}, we investigate the properties and microscopic structures of mixed phases for various types of first-order phase transitions in a unified manner, where the corresponding surface tensions are estimated adopting a thin-wall approximation~\cite{Avancini2010_PRC82-055807}.
In particular, based on the Thomas-Fermi approximation, we first fix the meson masses of the extended NJL model by reproducing the binding energies of finite nuclei. Adopting the spherical and cylindrical approximations for the WS cells, we then examine the possible formation of geometrical structures for three types of first-order phase transitions, and our findings are listed as follows.
\begin{itemize}
  \item Liquid-gas phase transition of nuclear matter: the EOS and microscopic structures of various types of nuclear pastas are fixed by searching for the configurations that minimize the energy of the system;
  \item Chiral phase transitions for $u$ and $d$ quarks: the geometrical structures do not emerge due to too small critical surface tensions $\sigma_\mathrm{c}$ according to Eq.~(\ref{eq:sigma_c});
  \item Deconfinement phase transition that entangled with chiral phase transition for $s$ quarks: the obtained surface tension is smaller than the critical value $\sigma<\sigma_\mathrm{c}$, suggesting the formation of geometrical structures for the mixed phases. It is found that the WS cell sizes $R_\mathrm{W}$ are generally small ($\lesssim 10$ fm), while the droplet, rod, slab, tube, and bubble phases emerge sequentially as density increases. Due to the partially restored chiral symmetry for $s$ quarks, there exist attractive interaction between SQM and hyperons, which may be important to identify the possible formation of strangelets in heavy-ion reactions.
\end{itemize}

Finally, the results presented here should be useful for us to understand the properties and structures of dense stellar matter throughout compact stars and in particular the matter state in the core regions, while more extensive calculations in a three-dimensional geometry with large box sizes are necessary for our future study~\cite{Okamoto2012_PLB713-284, Okamoto2013_PRC88-025801, Xia2021_PRC103-055812, Xia2022_PRD106-063020}.

\begin{acknowledgments}
This work was supported by the National Natural Science Foundation of China (Grant No. 12275234), the National SKA Program of China (Grant No. 2020SKA0120300), and  JSPS KAKENHI (Grant Nos. 20H04742 and 20K03951).
\end{acknowledgments}


%

\end{document}